\documentclass{article}

\usepackage{graphicx}
\PassOptionsToPackage{round}{natbib}
\usepackage[dblblindworkshop, final]{neurips_2025}
\usepackage{aas_macros}
\usepackage{bm}

\usepackage[utf8]{inputenc} 
\usepackage[T1]{fontenc}    
\usepackage{hyperref}       
\usepackage{url}            
\usepackage{booktabs}       
\usepackage{amsfonts}       
\usepackage{nicefrac}       
\usepackage{microtype}      
\usepackage{xcolor}         

\title{Characterizing Stellar Streams with Error-Aware Machine Learning}

\author{%
Alexandros Pratsos$^{1}$, \ Biprateep Dey$^{1,2}$\thanks{Corresponding author: Biprateep Dey (\texttt{biprateep@pitt.edu})}, \ Ting S.~Li$^{1}$ \\ \\
$^{1}$University of Toronto \quad $^{2}$Vector Institute\\
}

\begin{document}

\maketitle

\begin{abstract}
Stellar streams are thin, elongated collections of stars formed by gravitational disruption of orbiting star clusters or dwarf galaxies and are highly sensitive probes of the Milky Way's dark matter distribution and formation history. We present \texttt{SCREAM} (\textbf{S}tream \textbf{C}ha\textbf{R}acterization with \textbf{E}rror \textbf{A}ware \textbf{M}achine Learning), a weakly-supervised framework to identify member stars of stellar streams. Building on the \texttt{CATHODE} method originally developed for particle physics, \texttt{SCREAM} identifies streams as localized feature-space over-densities, avoiding rigid physical priors like assumed gravitational potentials or strict isochrone filtering. Crucially, \texttt{SCREAM} is the first machine learning (ML) framework in this domain to directly incorporate observational uncertainties into the neural network training objective. Using astrometric and photometric data from Gaia Data Release 3 and the Dark Energy Spectroscopic Instrument (DESI) Legacy imaging survey, we demonstrate our algorithm's performance on the prominent GD-1 stream. Validated against independent labels, \texttt{SCREAM} achieves an F1 score of 0.745, substantially outperforming existing ML methods in both precision and recall. Furthermore, \texttt{SCREAM} recovers the physically expected diffuse ``cocoon'' of GD-1 and faint main-sequence members that classical physics-based algorithms (e.g., \texttt{STREAMFINDER}) miss. Our results highlight the transformative potential of uncertainty-aware, weakly-supervised ML to uncover complex galactic structures.
\end{abstract}

\section{Introduction} \label{sec:intro}



Stellar streams are narrow, extended structures formed when gravitationally bound progenitors (e.g., globular clusters, dwarf galaxies) are disrupted by their host's gravitational field \citep{Johnston1998TDE}. The Milky Way hosts over a hundred known streams, typically containing $\mathcal{O}(100-1000)$ observable stars \citep{Mateu2023Galstreams}. Their physical narrowness and low velocity dispersions make them sensitive probes of dark matter interactions within the galactic halo \citep{LyndenBell1995Stream}.

Stream identification from large astronomical surveys has historically relied on both physics-based and data-driven methods \citep{Bonaca2025Streamreview}. While approaches like \texttt{STREAMFINDER} \citep{Malhan2018Streamfinder} are highly successful, they depend on assumed models of the Milky Way's gravitational potential and rigid stellar evolution priors. To avoid these dependencies, recent frameworks such as \texttt{VIA MACHINAE} \citep{Shih2022Viamachinae1,Shih2024Viamachinae2,Hallin2025Viamachinae3}, \citet{Pettee2024StreamCwola}, and \texttt{SKYCURTAINS} \citep{Sengupta2025Skycurtains} treat stream detection as weakly-supervised anomaly detection. They adapt particle physics techniques like \texttt{ANODE} \citep{Nachman2020Anode}, \texttt{CWoLA} \citep{Metodiev2017CWoLA}, and \texttt{CATHODE} \citep{Hallin2022Cathode} to model streams as localized over-densities in the feature space. However, these methods often ignore uncertainties in the input measurements, employ heavy post-processing (e.g., line-finding algorithms), and focus primarily on identifying new stream candidates over finding member stars of an already known stream (also called characterization).

We present \texttt{SCREAM} (\textbf{S}tream \textbf{C}ha\textbf{R}acterization with \textbf{E}rror \textbf{A}ware \textbf{M}achine learning), an anomaly detection framework designed to account for observational uncertainties in stellar stream searches and characterization. Building on \texttt{CATHODE}, \texttt{SCREAM} marginalizes over feature uncertainties within the loss function to more accurately capture the underlying distribution. Our approach requires minimal physical priors, accommodates arbitrary datasets, integrates uncertainties during training, and eliminates post-processing steps. This preserves anomaly score statistics and enables the discovery of members that rigid models might miss. While extensible to blind detection, this work focuses on characterizing known streams. We demonstrate \texttt{SCREAM} on the GD-1 stream \citep{Grillmair2006GD1}, validating predictions against independent reliable labels \citep{2026arXiv260420958J} to confirm the algorithm's ability to recover the true stellar stream distribution.

\section{Data and Methods} \label{methods}

\begin{figure*}
    \centering
    \includegraphics[width=\textwidth]{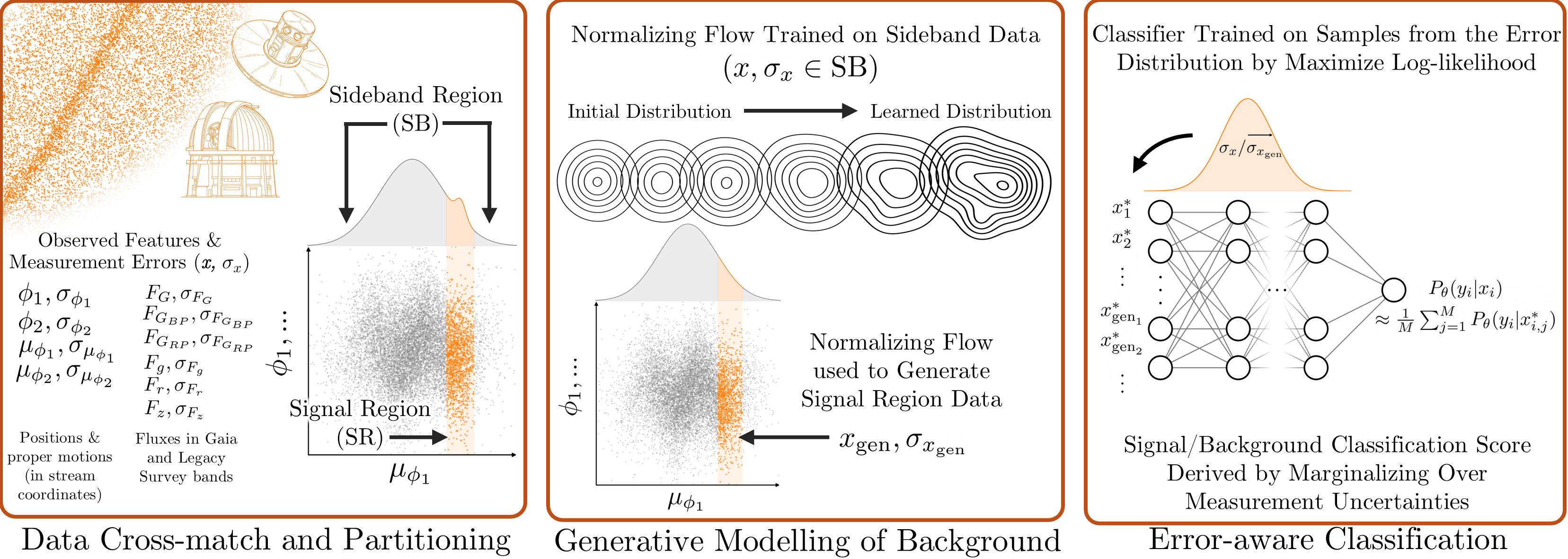}
    \caption{\small \textbf{Schematic representation of the \texttt{SCREAM} methodology}. \textbf{Left (Data Cross-match and Partitioning):} Kinematic and photometric features ($x$) and associated measurement uncertainties ($\sigma_x$), are compiled from Gaia and the DESI Legacy Survey. Data is partitioned into a Signal Region (SR) and a background-dominated Sideband Region (SB) based on along-stream proper motion ($\mu_{\phi_1}$). \textbf{Middle (Generative Modeling of Background):} A conditional Normalizing Flow trained on the SB data learns the joint distribution of background features and errors. It acts as an interpolator to generate realistic background data ($x_{\mathrm{gen}}, \sigma_{x_{\mathrm{gen}}}$) in the SR. \textbf{Right (Error-aware Classification):} A neural network classifier distinguishes true SR data from synthetic background. The training marginalizes over observational uncertainties by Monte Carlo sampling from the Gaussian error distributions, yielding a robust classification score.}
    \label{fig:workflow}
\end{figure*}

\paragraph{Data Sets Used.} We utilize astrometric and photometric data from Gaia Data Release 3 (DR3) \citep{Gaia2023DR3}, cross-matched with the DESI Legacy Imaging Survey DR9 \citep{Dey2019LegacySurvey} for additional photometric features. We query objects present in both catalogs within a $3^\circ$ padding of the GD-1 stream track defined in \texttt{galstreams} \citep{Mateu2023Galstreams}. Gaia provides sky positions and proper motions (angular velocities) transformed into a local, stream-aligned coordinate frame: $\phi_1$ aligns with the projected path, $\phi_2$ is the perpendicular transverse angle, and $\mu_{\phi_{1}}$ and $\mu_{\phi_{2}}$ denote their respective proper motions. To complement these kinematic data, we include broadband flux measurements ($F_{G}$, $F_{G_{BP}}$, $F_{G_{RP}}$, $F_g$, $F_r$, $F_z$), corrected for Milky Way dust extinction \citep{Schlegel1998Dustmap}. Finally, we explicitly incorporate the observational uncertainties (denoted as $\sigma$ with the corresponding subscript) for all features directly into our framework.

\paragraph{The \texttt{CATHODE} Framework.} We build our algorithm by extending \texttt{CATHODE} \citep{Hallin2022Cathode}, a weakly-supervised anomaly detection method originally developed to identify anomalous populations in high-dimensional particle physics data. We first partition the data into two non-random subsets, a signal region (SR) and its complement, the sideband region (SB), using the value of a single feature. We define SR using the along-stream proper motion, $\mu_{\phi_1}$, as streams are tightly kinematically constrained in this dimension. For GD-1, we define the SR as the 5th to 95th percentile range of $\mu_{\phi_1}$ from the \texttt{STREAMFINDER} catalog. Every observation falls into either the SR or SB based strictly on its $\mu_{\phi_1}$ value.

Next, we perform conditional density estimation of the background in the SB, assuming $P_{\mathrm{bg}}(\mathbf{x}, \sigma_{\mathbf{x}}|\mu_{\phi_1} \in \mathrm{SB}) \approx P_{\mathrm{data}}(\mathbf{x}, \sigma_{\mathbf{x}}|\mu_{\phi_1} \in \mathrm{SB})$. We implement this via conditional Normalizing Flows (NF) using \texttt{pzflow} \citep{Crenshaw2024pzflow} with Rational-Quadratic Rolling Spline Couplings \citep{Dinh2014Nice,Durkan2019SplineFlow}. We append measurement uncertainties to the NF's input space, jointly modeling features and errors so generated backgrounds possess realistic uncertainties. Leveraging the NF's interpolation capabilities, we generate background-like data within the SR. We fit a Kernel Density Estimate (KDE) to the $\mu_{\phi_1}$ distribution of the observed SR data, sample $\mu_{\phi_1} \sim P_{\mathrm{KDE}}(\mu_{\phi_1})$, and apply the inverse NF transformation to generate synthetic background features ($\mathbf{x}_{\mathrm{gen}}, \sigma_{\mathbf{x_{\mathrm{gen}}}}$). To prevent class imbalance, we generate the same number of samples as the true SR observations.

A neural network then classifies observed versus generated background data in the SR. Since the generated data models the background, the classifier cannot differentiate the two without a signal. Thus, any learned separation is directly attributable to the stellar stream signal. Although these class probabilities are miscalibrated, they preserve a monotonic relationship with the true underlying signal probabilities \citep{Metodiev2017CWoLA}.

\paragraph{Extending \texttt{CATHODE} to Incorporate Measurement Uncertainties.}
 When training a standard classifier, the objective is to maximize the likelihood of the true labels given deterministic inputs. In \texttt{SCREAM}, we treat the observed features $\mathbf{x}_i$ for an object $i$ as a single realization drawn from a distribution centered around the true, unobserved features $\mathbf{x}_i^{\mathrm{true}}$. The likelihood for a single observation is:
\begin{equation}
    P_\theta(y_i | \mathbf{x}_i) = \int P_\theta(y_i | \mathbf{x}_i^{\mathrm{true}}) P(\mathbf{x}_i^{\mathrm{true}} | \mathbf{x}_i) \, d\mathbf{x}_i^{\mathrm{true}}.
\end{equation}

Using Bayes' theorem and assuming a uniform prior over the true features, we obtain:
\begin{equation}
    P_\theta(y_i | \mathbf{x}_i) \propto \int P_\theta(y_i | \mathbf{x}_i^{\mathrm{true}}) P(\mathbf{x}_i | \mathbf{x}_i^{\mathrm{true}}) \, d\mathbf{x}_i^{\mathrm{true}},
\end{equation}
where $P(\mathbf{x}_i | \mathbf{x}_i^{\mathrm{true}})$ is the distribution from which our observed features are drawn, modeled here as an uncorrelated multidimensional Gaussian. Exploiting the symmetry of the Gaussian distribution, we approximate this analytically intractable integral using Monte Carlo (MC) integration.  We draw $M$ feature samples, $\mathbf{x}_{i,j}^{*}$, from the noise distribution centered on the observation $\mathbf{x}_i$ with a diagonal covariance matrix defined by $\sigma_{\mathbf{x}_i}$. The integral is then approximated by the summation:
\begin{equation}
    P_\theta(y_i | \mathbf{x}_i) \approx \frac{1}{M} \sum_{j=1}^M P_\theta(y_i | \mathbf{x}_{i,j}^{*}).
\end{equation}

For a binary classification problem, $P_\theta(y_i | \mathbf{x}_{i,j}^{*})$ follows a Bernoulli distribution. Therefore, the approximated negative log-likelihood (loss) for a single observation is:
\begin{equation}
    \mathcal{L}_{\mathrm{MC}} = -\ln\left(\frac{1}{M}\sum_{j=1}^M p_{i,j}^{y_i}(1-p_{i,j})^{1-y_i}\right).
\end{equation}
where, $p_{i,j}$ is the predicted class probability. This loss is averaged over the training set and minimized to find the optimal model parameters. Though novel for this domain, a similar approach to incorporate input measurement uncertainties have been used in other contexts in previous work such as \citet{Wang2023SBI} and \citet{Sun2023Zephyr}. A schematic overview of the whole algorithm is shown in Figure~\ref{fig:workflow}.

Our classification network is a Multi-Layer Perceptron \citep{Rosenblatt1958Perceptron} with GELU non-linearities \citep{Hendrycks2016Gelu}, optimized via Adam \citep{Kingma2015Adam}. Full architectural details are in our code repository\footnote{\url{https://github.com/pratsosa/SCREAM}}. To increase stochastic regularization, we set $M=10$ during training. During inference, we draw $M=100$ samples and average them to obtain the classification score for each object. Predictions for the entire data set are generated by $k=5$-fold cross-validation.

\section{Results and Discussion} \label{sec:results}

\begin{figure*}[t]
    \centering
    \includegraphics[width=\textwidth]{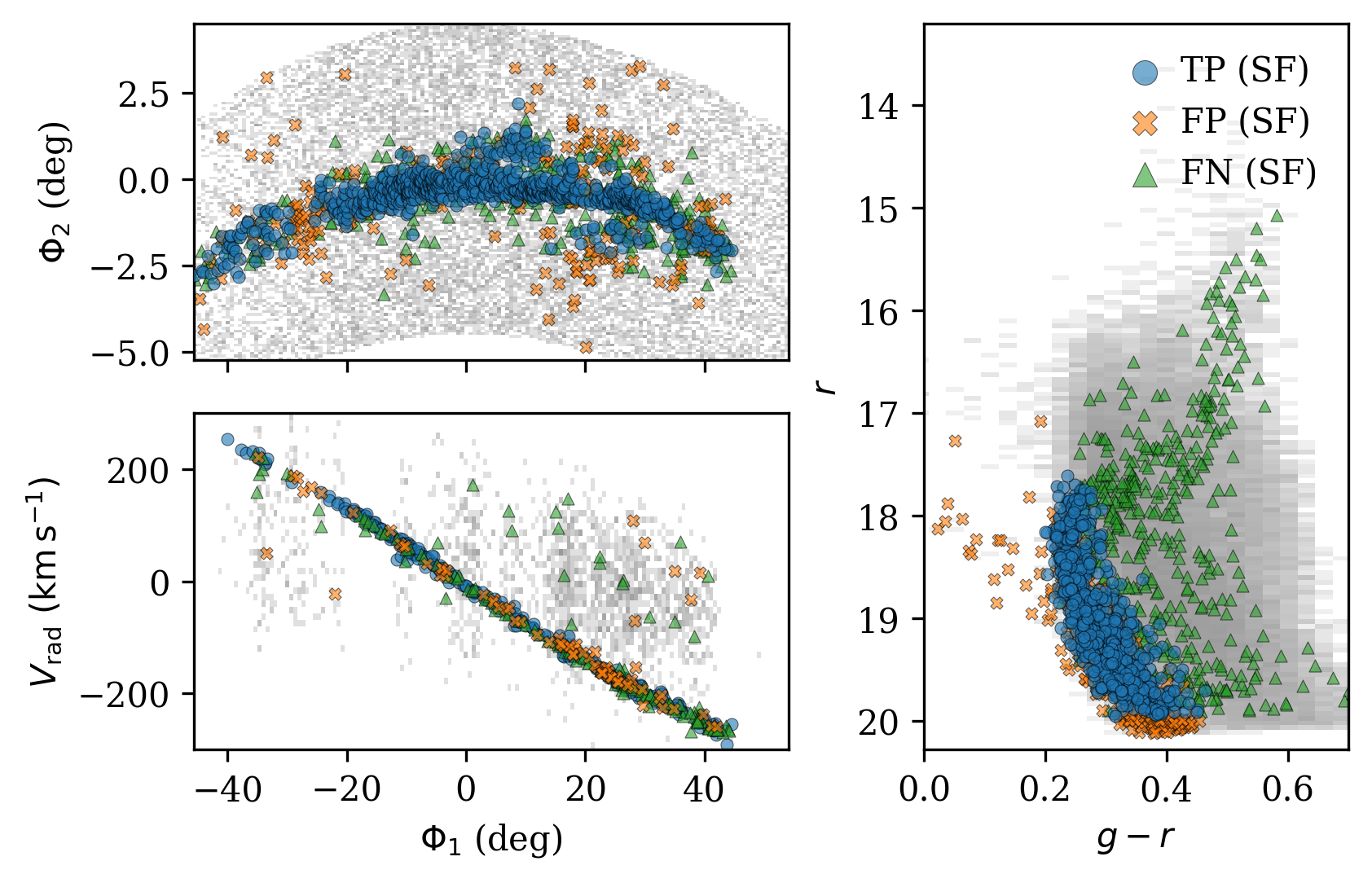}
    \caption{\textbf{Visualization of SCREAM performance relative to the STREAMFINDER (SF) baseline.} The background distribution is shown in gray. \textbf{Top Left ($\phi_2$ vs. $\phi_1$):} \texttt{SCREAM} recovers stars along the full spatial extent of GD-1. Many apparent False Positives (orange crosses) located off the narrow central track are confirmed members via independent radial velocity labels. \textbf{Right ($r$ vs. $g-r$):} Predictions trace the dense main sequence. While missing brighter RGB stars ($r < 18.5$), \texttt{SCREAM} recovers more faint main sequence stars than SF. \textbf{Bottom Left ($V_{\mathrm{rad}}$ vs. $\phi_1$):} \texttt{SCREAM} False Positives align with the true kinematic track, demonstrating high precision identifying genuine members SF missed. False Negatives (green triangles) indicate recall can be improved.}
    \label{fig:result}
\end{figure*}

We evaluate the performance of \texttt{SCREAM} against two distinct reference datasets: \texttt{STREAMFINDER} (SF), serving as a proxy for an algorithmic baseline and the highly reliable labels derived from DESI radial velocity measurements from \citet{2026arXiv260420958J}. 

 At a 0.878 confidence threshold, \texttt{SCREAM} achieves an F1 score of 0.66 (Precision: 0.69, Recall: 0.62) against SF labels. Evaluated against true spectroscopic labels, performance improves to an F1 of 0.745 (Accuracy: 0.959, Precision: 0.871, Recall: 0.650). This indicates our approach captures the true physical distribution more accurately than the SF baseline. While differing datasets complicate direct comparisons with other machine learning-based methods, \texttt{SCREAM} outperforms algorithms like \texttt{SKYCURTAINS} \citep{Sengupta2025Skycurtains} by 12 percentage points in precision and 28 in recall, while also accounting for input measurement uncertainties in the classifications.

Figure~\ref{fig:result} visualizes the algorithm performance with respect to \texttt{STREAMFINDER}. Spatially (top-left), \texttt{SCREAM} uniquely recovers the stream's diffuse ``cocoon'', a physically expected feature often missed by SF's rigid priors. Thus, many apparent False Positives relative to SF (orange crosses) are genuine members. Photometrically (right), minimizing physical priors allows \texttt{SCREAM} to recover fainter member stars that SF misses. However, the model struggles with brighter Red Giant Branch (RGB) stars as their lower feature-space density causes the anomaly detector to miss them. Kinematically (bottom-left), apparent False Positives lie almost entirely on the true kinematic track, reinforcing high true precision, though False Negatives on the track highlight room for recall improvement.

To our knowledge, \texttt{SCREAM} is among the first ML-based stellar stream identification and characterization frameworks to incorporate observational uncertainties directly into the training objective. Because SF labels were never used during training, \texttt{SCREAM} is not restricted by previous assumptions. Future work will address current limitations. We will exploit the model's probabilistic nature to handle missing data and feature scaling and density estimation improvements will enable the detection of low-density RGB stars. A thorough vetting of our algorithm on mock data will also be performed in a future work. Finally, implementing a scanning signal-region window and sampling from full, non-diagonal error covariance matrices will enable blind, all-sky searches and improved classification across multiple Milky Way streams.

\begin{ack}
The authors acknowledge the support of the Data Sciences Institute, University of Toronto via the 2025 Summer Undergraduate Data Science (SUDS) Opportunities Program. B. Dey is a postdoctoral fellow at the University of Toronto in the Eric and Wendy Schmidt AI in Science Postdoctoral Fellowship Program, a program of Schmidt Sciences.

The authors advocate for the judicious and ethical use of artificial intelligence in scientific discourse. In accordance with the \href{https://journals.aps.org/authors/ai-based-writing-tools}{American Physical Society’s guidelines} on appropriate use of AI-based writing tools, we confirm that AI assistance was utilized in the preparation of this statement solely to polish, condense, and edit our original ideas. The authors retain full responsibility for the accuracy, integrity, and content of this document.

This research used data obtained with the Dark Energy Spectroscopic Instrument (DESI). DESI construction and operations is managed by the Lawrence Berkeley National Laboratory. This material is based upon work supported by the U.S. Department of Energy, Office of Science, Office of High-Energy Physics, under Contract No. DE–AC02–05CH11231, and by the National Energy Research Scientific Computing Center, a DOE Office of Science User Facility under the same contract. Additional support for DESI was provided by the U.S. National Science Foundation (NSF), Division of Astronomical Sciences under Contract No. AST-0950945 to the NSF’s National Optical-Infrared Astronomy Research Laboratory; the Science and Technology Facilities Council of the United Kingdom; the Gordon and Betty Moore Foundation; the Heising-Simons Foundation; the French Alternative Energies and Atomic Energy Commission (CEA); the National Council of Humanities, Science and Technology of Mexico (CONAHCYT); the Ministry of Science and Innovation of Spain (MICINN), and by the DESI Member Institutions: www.desi.lbl.gov/collaborating-institutions. The DESI collaboration is honored to be permitted to conduct scientific research on I’oligam Du’ag (Kitt Peak), a mountain with particular significance to the Tohono O’odham Nation. Any opinions, findings, and conclusions or recommendations expressed in this material are those of the author(s) and do not necessarily reflect the views of the U.S. National Science Foundation, the U.S. Department of Energy, or any of the listed funding agencies.

The Legacy Surveys consist of three individual and complementary projects: the Dark Energy Camera Legacy Survey (DECaLS; Proposal ID \#2014B-0404; PIs: David Schlegel and Arjun Dey), the Beijing-Arizona Sky Survey (BASS; NOAO Prop. ID \#2015A-0801; PIs: Zhou Xu and Xiaohui Fan), and the Mayall z-band Legacy Survey (MzLS; Prop. ID \#2016A-0453; PI: Arjun Dey). DECaLS, BASS and MzLS together include data obtained, respectively, at the Blanco telescope, Cerro Tololo Inter-American Observatory, NSF’s NOIRLab; the Bok telescope, Steward Observatory, University of Arizona; and the Mayall telescope, Kitt Peak National Observatory, NOIRLab. Pipeline processing and analyses of the data were supported by NOIRLab and the Lawrence Berkeley National Laboratory (LBNL). The Legacy Surveys project is honored to be permitted to conduct astronomical research on Iolkam Du’ag (Kitt Peak), a mountain with particular significance to the Tohono O’odham Nation.

NOIRLab is operated by the Association of Universities for Research in Astronomy (AURA) under a cooperative agreement with the National Science Foundation. LBNL is managed by the Regents of the University of California under contract to the U.S. Department of Energy.

This project used data obtained with the Dark Energy Camera (DECam), which was constructed by the Dark Energy Survey (DES) collaboration. Funding for the DES Projects has been provided by the U.S. Department of Energy, the U.S. National Science Foundation, the Ministry of Science and Education of Spain, the Science and Technology Facilities Council of the United Kingdom, the Higher Education Funding Council for England, the National Center for Supercomputing Applications at the University of Illinois at Urbana-Champaign, the Kavli Institute of Cosmological Physics at the University of Chicago, Center for Cosmology and Astro-Particle Physics at the Ohio State University, the Mitchell Institute for Fundamental Physics and Astronomy at Texas A\&M University, Financiadora de Estudos e Projetos, Fundacao Carlos Chagas Filho de Amparo, Financiadora de Estudos e Projetos, Fundacao Carlos Chagas Filho de Amparo a Pesquisa do Estado do Rio de Janeiro, Conselho Nacional de Desenvolvimento Cientifico e Tecnologico and the Ministerio da Ciencia, Tecnologia e Inovacao, the Deutsche Forschungsgemeinschaft and the Collaborating Institutions in the Dark Energy Survey. The Collaborating Institutions are Argonne National Laboratory, the University of California at Santa Cruz, the University of Cambridge, Centro de Investigaciones Energeticas, Medioambientales y Tecnologicas-Madrid, the University of Chicago, University College London, the DES-Brazil Consortium, the University of Edinburgh, the Eidgenossische Technische Hochschule (ETH) Zurich, Fermi National Accelerator Laboratory, the University of Illinois at Urbana-Champaign, the Institut de Ciencies de l’Espai (IEEC/CSIC), the Institut de Fisica d’Altes Energies, Lawrence Berkeley National Laboratory, the Ludwig Maximilians Universitat Munchen and the associated Excellence Cluster Universe, the University of Michigan, NSF’s NOIRLab, the University of Nottingham, the Ohio State University, the University of Pennsylvania, the University of Portsmouth, SLAC National Accelerator Laboratory, Stanford University, the University of Sussex, and Texas A\&M University.

BASS is a key project of the Telescope Access Program (TAP), which has been funded by the National Astronomical Observatories of China, the Chinese Academy of Sciences (the Strategic Priority Research Program “The Emergence of Cosmological Structures” Grant \# XDB09000000), and the Special Fund for Astronomy from the Ministry of Finance. The BASS is also supported by the External Cooperation Program of Chinese Academy of Sciences (Grant \# 114A11KYSB20160057), and Chinese National Natural Science Foundation (Grant \# 12120101003, \# 11433005).

The Legacy Survey team makes use of data products from the Near-Earth Object Wide-field Infrared Survey Explorer (NEOWISE), which is a project of the Jet Propulsion Laboratory/California Institute of Technology. NEOWISE is funded by the National Aeronautics and Space Administration.

The Legacy Surveys imaging of the DESI footprint is supported by the Director, Office of Science, Office of High Energy Physics of the U.S. Department of Energy under Contract No. DE-AC02-05CH1123, by the National Energy Research Scientific Computing Center, a DOE Office of Science User Facility under the same contract; and by the U.S. National Science Foundation, Division of Astronomical Sciences under Contract No. AST-0950945 to NOAO.

This work has made use of data from the European Space Agency (ESA) mission
{\it Gaia} (\url{https://www.cosmos.esa.int/gaia}), processed by the {\it Gaia}
Data Processing and Analysis Consortium (DPAC,
\url{https://www.cosmos.esa.int/web/gaia/dpac/consortium}). Funding for the DPAC
has been provided by national institutions, in particular the institutions
participating in the {\it Gaia} Multilateral Agreement.
\end{ack}


\bibliographystyle{plainnat}
\bibliography{references}

@ARTICLE{2026arXiv260420958J,
       author = {{Jarvis}, Emma and {Li}, Ting S. and {Koposov}, Sergey E. and {Carlberg}, Raymond G. and {Valluri}, Monica and {Mohammed}, Nasser and {Aguilar}, J. and {Ahlen}, S. and {Allende Prieto}, Carlos and {Beraldo e Silva}, Leandro and {Bianchi}, D. and {Brooks}, D. and {Bystr{\"o}m}, Amanda and {Claybaugh}, T. and {Cooper}, A.~P. and {Cuceu}, A. and {de la Macorra}, A. and {Dey}, Arjun and {Dey}, Biprateep and {Doel}, P. and {Forero-Romero}, J.~E. and {Gazta{\~n}aga}, E. and {Gnedin}, Oleg Y. and {Gontcho}, Satya Gontcho A and {Gutierrez}, G. and {Honscheid}, K. and {Joyce}, R. and {Kehoe}, R. and {Kisner}, T. and {Kizhuprakkat}, Namitha and {Kremin}, A. and {Lambert}, Mika and {Landriau}, M. and {Le Guillou}, L. and {Medina}, Gustavo E. and {Meisner}, A. and {Miquel}, R. and {Nadathur}, S. and {Najita}, Joan and {Palanque-Delabrouille}, N. and {Percival}, W.~J. and {Prada}, F. and {P{\'e}rez-R{\`a}fols}, I. and {Qiu}, Tian and {Riley}, Alexander H. and {Rockosi}, Constance M. and {Rossi}, G. and {Sanchez}, E. and {Sandford}, Nathan and {Schlafly}, E.~F. and {Schlegel}, D. and {Silber}, J. and {Sprayberry}, D. and {Tarl{\'e}}, G. and {Weaver}, B.~A. and {Zhou}, R. and {Zou}, H.},
        title = "{Characterizing the GD-1 Stream with DESI DR2 Data: Thin Stream and Hot Cocoon}",
      journal = {arXiv e-prints},
         year = 2026,
        month = apr,
          eid = {arXiv:2604.20958},
        pages = {arXiv:2604.20958},
          doi = {10.48550/arXiv.2604.20958},
archivePrefix = {arXiv},
       eprint = {2604.20958},
 primaryClass = {astro-ph.GA},
       adsurl = {https://ui.adsabs.harvard.edu/abs/2026arXiv260420958J}
}

@article{Hendrycks2016Gelu,
  title={Gaussian error linear units (gelus)},
  author={Hendrycks, Dan and Gimpel, Kevin},
  journal={arXiv preprint arXiv:1606.08415},
  year={2016}
}

@article{Rosenblatt1958Perceptron,
  title={The perceptron: a probabilistic model for information storage and organization in the brain.},
  author={Rosenblatt, Frank},
  journal={Psychological review},
  volume={65},
  number={6},
  pages={386},
  year={1958},
  publisher={American Psychological Association}
}

@inproceedings{Kingma2015Adam,
  author    = {Kingma, Diederik P. and Ba, Jimmy},
  title     = {Adam: A Method for Stochastic Optimization},
  booktitle = {International Conference on Learning Representations (ICLR)},
  year      = {2015},
  url       = {https://arxiv.org/abs/1412.6980}
}

@ARTICLE{Wang2023SBI,
       author = {{Wang}, Bingjie and {Leja}, Joel and {Villar}, V. Ashley and {Speagle}, Joshua S.},
        title = "{SBI$^{++}$: Flexible, Ultra-fast Likelihood-free Inference Customized for Astronomical Applications}",
      journal = {\apjl},
         year = 2023,
        month = jul,
       volume = {952},
       number = {1},
          eid = {L10},
        pages = {L10},
          doi = {10.3847/2041-8213/ace361},
archivePrefix = {arXiv},
       eprint = {2304.05281},
 primaryClass = {astro-ph.IM},
       adsurl = {https://ui.adsabs.harvard.edu/abs/2023ApJ...952L..10W}
}

@ARTICLE{Sun2023Zephyr,
       author = {{Sun}, Zechang and {Speagle}, Joshua S. and {Huang}, Song and {Ting}, Yuan-Sen and {Cai}, Zheng},
        title = "{Zephyr : Stitching Heterogeneous Training Data with Normalizing Flows for Photometric Redshift Inference}",
      journal = {arXiv e-prints},
         year = 2023,
        month = oct,
          eid = {arXiv:2310.20125},
        pages = {arXiv:2310.20125},
          doi = {10.48550/arXiv.2310.20125},
archivePrefix = {arXiv},
       eprint = {2310.20125},
 primaryClass = {astro-ph.IM},
       adsurl = {https://ui.adsabs.harvard.edu/abs/2023arXiv231020125S}
}

@ARTICLE{Durkan2019SplineFlow,
       author = {{Durkan}, Conor and {Bekasov}, Artur and {Murray}, Iain and {Papamakarios}, George},
        title = "{Neural Spline Flows}",
      journal = {arXiv e-prints},
         year = 2019,
        month = jun,
          eid = {arXiv:1906.04032},
        pages = {arXiv:1906.04032},
          doi = {10.48550/arXiv.1906.04032},
archivePrefix = {arXiv},
       eprint = {1906.04032},
 primaryClass = {stat.ML},
       adsurl = {https://ui.adsabs.harvard.edu/abs/2019arXiv190604032D}
}

@ARTICLE{Dinh2014Nice,
       author = {{Dinh}, Laurent and {Krueger}, David and {Bengio}, Yoshua},
        title = "{NICE: Non-linear Independent Components Estimation}",
      journal = {arXiv e-prints},
         year = 2014,
        month = oct,
          eid = {arXiv:1410.8516},
        pages = {arXiv:1410.8516},
          doi = {10.48550/arXiv.1410.8516},
archivePrefix = {arXiv},
       eprint = {1410.8516},
 primaryClass = {cs.LG},
       adsurl = {https://ui.adsabs.harvard.edu/abs/2014arXiv1410.8516D}
}

@ARTICLE{Crenshaw2024pzflow,
       author = {{Crenshaw}, John Franklin and {Kalmbach}, J. Bryce and {Gagliano}, Alexander and {Yan}, Ziang and {Connolly}, Andrew J. and {Malz}, Alex I. and {Schmidt}, Samuel J. and {The LSST Dark Energy Science Collaboration}},
        title = "{Probabilistic Forward Modeling of Galaxy Catalogs with Normalizing Flows}",
      journal = {\aj},
         year = 2024,
        month = aug,
       volume = {168},
       number = {2},
          eid = {80},
        pages = {80},
          doi = {10.3847/1538-3881/ad54bf},
archivePrefix = {arXiv},
       eprint = {2405.04740},
 primaryClass = {astro-ph.IM},
       adsurl = {https://ui.adsabs.harvard.edu/abs/2024AJ....168...80C}
}

@ARTICLE{Schlegel1998Dustmap,
       author = {{Schlegel}, David J. and {Finkbeiner}, Douglas P. and {Davis}, Marc},
        title = "{Maps of Dust Infrared Emission for Use in Estimation of Reddening and Cosmic Microwave Background Radiation Foregrounds}",
      journal = {\apj},
         year = 1998,
        month = jun,
       volume = {500},
       number = {2},
        pages = {525-553},
          doi = {10.1086/305772},
archivePrefix = {arXiv},
       eprint = {astro-ph/9710327},
 primaryClass = {astro-ph},
       adsurl = {https://ui.adsabs.harvard.edu/abs/1998ApJ...500..525S}
}

@ARTICLE{Dey2019LegacySurvey,
       author = {{Dey}, Arjun and {Schlegel}, David J. and {Lang}, Dustin and {Blum}, Robert and {Burleigh}, Kaylan and {Fan}, Xiaohui and {Findlay}, Joseph R. and {Finkbeiner}, Doug and {Herrera}, David and {Juneau}, St{\'e}phanie and {Landriau}, Martin and {Levi}, Michael and {McGreer}, Ian and {Meisner}, Aaron and {Myers}, Adam D. and {Moustakas}, John and {Nugent}, Peter and {Patej}, Anna and {Schlafly}, Edward F. and {Walker}, Alistair R. and {Valdes}, Francisco and {Weaver}, Benjamin A. and {Y{\`e}che}, Christophe and {Zou}, Hu and {Zhou}, Xu and {Abareshi}, Behzad and {Abbott}, T.~M.~C. and {Abolfathi}, Bela and {Aguilera}, C. and {Alam}, Shadab and {Allen}, Lori and {Alvarez}, A. and {Annis}, James and {Ansarinejad}, Behzad and {Aubert}, Marie and {Beechert}, Jacqueline and {Bell}, Eric F. and {BenZvi}, Segev Y. and {Beutler}, Florian and {Bielby}, Richard M. and {Bolton}, Adam S. and {Brice{\~n}o}, C{\'e}sar and {Buckley-Geer}, Elizabeth J. and {Butler}, Karen and {Calamida}, Annalisa and {Carlberg}, Raymond G. and {Carter}, Paul and {Casas}, Ricard and {Castander}, Francisco J. and {Choi}, Yumi and {Comparat}, Johan and {Cukanovaite}, Elena and {Delubac}, Timoth{\'e}e and {DeVries}, Kaitlin and {Dey}, Sharmila and {Dhungana}, Govinda and {Dickinson}, Mark and {Ding}, Zhejie and {Donaldson}, John B. and {Duan}, Yutong and {Duckworth}, Christopher J. and {Eftekharzadeh}, Sarah and {Eisenstein}, Daniel J. and {Etourneau}, Thomas and {Fagrelius}, Parker A. and {Farihi}, Jay and {Fitzpatrick}, Mike and {Font-Ribera}, Andreu and {Fulmer}, Leah and {G{\"a}nsicke}, Boris T. and {Gaztanaga}, Enrique and {George}, Koshy and {Gerdes}, David W. and {Gontcho}, Satya Gontcho A. and {Gorgoni}, Claudio and {Green}, Gregory and {Guy}, Julien and {Harmer}, Diane and {Hernandez}, M. and {Honscheid}, Klaus and {Huang}, Lijuan Wendy and {James}, David J. and {Jannuzi}, Buell T. and {Jiang}, Linhua and {Joyce}, Richard and {Karcher}, Armin and {Karkar}, Sonia and {Kehoe}, Robert and {Kneib}, Jean-Paul and {Kueter-Young}, Andrea and {Lan}, Ting-Wen and {Lauer}, Tod R. and {Le Guillou}, Laurent and {Le Van Suu}, Auguste and {Lee}, Jae Hyeon and {Lesser}, Michael and {Perreault Levasseur}, Laurence and {Li}, Ting S. and {Mann}, Justin L. and {Marshall}, Robert and {Mart{\'\i}nez-V{\'a}zquez}, C.~E. and {Martini}, Paul and {du Mas des Bourboux}, H{\'e}lion and {McManus}, Sean and {Meier}, Tobias Gabriel and {M{\'e}nard}, Brice and {Metcalfe}, Nigel and {Mu{\~n}oz-Guti{\'e}rrez}, Andrea and {Najita}, Joan and {Napier}, Kevin and {Narayan}, Gautham and {Newman}, Jeffrey A. and {Nie}, Jundan and {Nord}, Brian and {Norman}, Dara J. and {Olsen}, Knut A.~G. and {Paat}, Anthony and {Palanque-Delabrouille}, Nathalie and {Peng}, Xiyan and {Poppett}, Claire L. and {Poremba}, Megan R. and {Prakash}, Abhishek and {Rabinowitz}, David and {Raichoor}, Anand and {Rezaie}, Mehdi and {Robertson}, A.~N. and {Roe}, Natalie A. and {Ross}, Ashley J. and {Ross}, Nicholas P. and {Rudnick}, Gregory and {Safonova}, Sasha and {Saha}, Abhijit and {S{\'a}nchez}, F. Javier and {Savary}, Elodie and {Schweiker}, Heidi and {Scott}, Adam and {Seo}, Hee-Jong and {Shan}, Huanyuan and {Silva}, David R. and {Slepian}, Zachary and {Soto}, Christian and {Sprayberry}, David and {Staten}, Ryan and {Stillman}, Coley M. and {Stupak}, Robert J. and {Summers}, David L. and {Sien Tie}, Suk and {Tirado}, H. and {Vargas-Maga{\~n}a}, Mariana and {Vivas}, A. Katherina and {Wechsler}, Risa H. and {Williams}, Doug and {Yang}, Jinyi and {Yang}, Qian and {Yapici}, Tolga and {Zaritsky}, Dennis and {Zenteno}, A. and {Zhang}, Kai and {Zhang}, Tianmeng and {Zhou}, Rongpu and {Zhou}, Zhimin},
        title = "{Overview of the DESI Legacy Imaging Surveys}",
      journal = {\aj},
         year = 2019,
        month = may,
       volume = {157},
       number = {5},
          eid = {168},
        pages = {168},
          doi = {10.3847/1538-3881/ab089d},
archivePrefix = {arXiv},
       eprint = {1804.08657},
 primaryClass = {astro-ph.IM},
       adsurl = {https://ui.adsabs.harvard.edu/abs/2019AJ....157..168D}
}

@ARTICLE{Gaia2023DR3,
       author = {{Gaia Collaboration} and {Vallenari}, A. and {Brown}, A.~G.~A. and {Prusti}, T. and {de Bruijne}, J.~H.~J. and {Arenou}, F. and {Babusiaux}, C. and {Biermann}, M. and {Creevey}, O.~L. and {Ducourant}, C. and {Evans}, D.~W. and {Eyer}, L. and {Guerra}, R. and {Hutton}, A. and {Jordi}, C. and {Klioner}, S.~A. and {Lammers}, U.~L. and {Lindegren}, L. and {Luri}, X. and {Mignard}, F. and {Panem}, C. and {Pourbaix}, D. and {Randich}, S. and {Sartoretti}, P. and {Soubiran}, C. and {Tanga}, P. and {Walton}, N.~A. and {Bailer-Jones}, C.~A.~L. and {Bastian}, U. and {Drimmel}, R. and {Jansen}, F. and {Katz}, D. and {Lattanzi}, M.~G. and {van Leeuwen}, F. and {Bakker}, J. and {Cacciari}, C. and {Casta{\~n}eda}, J. and {De Angeli}, F. and {Fabricius}, C. and {Fouesneau}, M. and {Fr{\'e}mat}, Y. and {Galluccio}, L. and {Guerrier}, A. and {Heiter}, U. and {Masana}, E. and {Messineo}, R. and {Mowlavi}, N. and {Nicolas}, C. and {Nienartowicz}, K. and {Pailler}, F. and {Panuzzo}, P. and {Riclet}, F. and {Roux}, W. and {Seabroke}, G.~M. and {Sordo}, R. and {Th{\'e}venin}, F. and {Gracia-Abril}, G. and {Portell}, J. and {Teyssier}, D. and {Altmann}, M. and {Andrae}, R. and {Audard}, M. and {Bellas-Velidis}, I. and {Benson}, K. and {Berthier}, J. and {Blomme}, R. and {Burgess}, P.~W. and {Busonero}, D. and {Busso}, G. and {C{\'a}novas}, H. and {Carry}, B. and {Cellino}, A. and {Cheek}, N. and {Clementini}, G. and {Damerdji}, Y. and {Davidson}, M. and {de Teodoro}, P. and {Nu{\~n}ez Campos}, M. and {Delchambre}, L. and {Dell'Oro}, A. and {Esquej}, P. and {Fern{\'a}ndez-Hern{\'a}ndez}, J. and {Fraile}, E. and {Garabato}, D. and {Garc{\'\i}a-Lario}, P. and {Gosset}, E. and {Haigron}, R. and {Halbwachs}, J.-L. and {Hambly}, N.~C. and {Harrison}, D.~L. and {Hern{\'a}ndez}, J. and {Hestroffer}, D. and {Hodgkin}, S.~T. and {Holl}, B. and {Jan{\ss}en}, K. and {Jevardat de Fombelle}, G. and {Jordan}, S. and {Krone-Martins}, A. and {Lanzafame}, A.~C. and {L{\"o}ffler}, W. and {Marchal}, O. and {Marrese}, P.~M. and {Moitinho}, A. and {Muinonen}, K. and {Osborne}, P. and {Pancino}, E. and {Pauwels}, T. and {Recio-Blanco}, A. and {Reyl{\'e}}, C. and {Riello}, M. and {Rimoldini}, L. and {Roegiers}, T. and {Rybizki}, J. and {Sarro}, L.~M. and {Siopis}, C. and {Smith}, M. and {Sozzetti}, A. and {Utrilla}, E. and {van Leeuwen}, M. and {Abbas}, U. and {{\'A}brah{\'a}m}, P. and {Abreu Aramburu}, A. and {Aerts}, C. and {Aguado}, J.~J. and {Ajaj}, M. and {Aldea-Montero}, F. and {Altavilla}, G. and {{\'A}lvarez}, M.~A. and {Alves}, J. and {Anders}, F. and {Anderson}, R.~I. and {Anglada Varela}, E. and {Antoja}, T. and {Baines}, D. and {Baker}, S.~G. and {Balaguer-N{\'u}{\~n}ez}, L. and {Balbinot}, E. and {Balog}, Z. and {Barache}, C. and {Barbato}, D. and {Barros}, M. and {Barstow}, M.~A. and {Bartolom{\'e}}, S. and {Bassilana}, J.-L. and {Bauchet}, N. and {Becciani}, U. and {Bellazzini}, M. and {Berihuete}, A. and {Bernet}, M. and {Bertone}, S. and {Bianchi}, L. and {Binnenfeld}, A. and {Blanco-Cuaresma}, S. and {Blazere}, A. and {Boch}, T. and {Bombrun}, A. and {Bossini}, D. and {Bouquillon}, S. and {Bragaglia}, A. and {Bramante}, L. and {Breedt}, E. and {Bressan}, A. and {Brouillet}, N. and {Brugaletta}, E. and {Bucciarelli}, B. and {Burlacu}, A. and {Butkevich}, A.~G. and {Buzzi}, R. and {Caffau}, E. and {Cancelliere}, R. and {Cantat-Gaudin}, T. and {Carballo}, R. and {Carlucci}, T. and {Carnerero}, M.~I. and {Carrasco}, J.~M. and {Casamiquela}, L. and {Castellani}, M. and {Castro-Ginard}, A. and {Chaoul}, L. and {Charlot}, P. and {Chemin}, L. and {Chiaramida}, V. and {Chiavassa}, A. and {Chornay}, N. and {Comoretto}, G. and {Contursi}, G. and {Cooper}, W.~J. and {Cornez}, T. and {Cowell}, S. and {Crifo}, F. and {Cropper}, M. and {Crosta}, M. and {Crowley}, C. and {Dafonte}, C. and {Dapergolas}, A. and {David}, M. and {David}, P. and {de Laverny}, P. and {De Luise}, F. and {De March}, R.},
        title = "{Gaia Data Release 3. Summary of the content and survey properties}",
      journal = {\aap},
         year = 2023,
        month = jun,
       volume = {674},
          eid = {A1},
        pages = {A1},
          doi = {10.1051/0004-6361/202243940},
archivePrefix = {arXiv},
       eprint = {2208.00211},
 primaryClass = {astro-ph.GA},
       adsurl = {https://ui.adsabs.harvard.edu/abs/2023A&A...674A...1G}
}

@ARTICLE{Grillmair2006GD1,
       author = {{Grillmair}, C.~J. and {Dionatos}, O.},
        title = "{Detection of a 63{\textdegree} Cold Stellar Stream in the Sloan Digital Sky Survey}",
      journal = {\apjl},
         year = 2006,
        month = may,
       volume = {643},
       number = {1},
        pages = {L17-L20},
          doi = {10.1086/505111},
archivePrefix = {arXiv},
       eprint = {astro-ph/0604332},
 primaryClass = {astro-ph},
       adsurl = {https://ui.adsabs.harvard.edu/abs/2006ApJ...643L..17G}
}

@ARTICLE{Hallin2022Cathode,
       author = {{Hallin}, Anna and {Isaacson}, Joshua and {Kasieczka}, Gregor and {Krause}, Claudius and {Nachman}, Benjamin and {Quadfasel}, Tobias and {Schlaffer}, Matthias and {Shih}, David and {Sommerhalder}, Manuel},
        title = "{Classifying anomalies through outer density estimation}",
      journal = {\prd},
         year = 2022,
        month = sep,
       volume = {106},
       number = {5},
          eid = {055006},
        pages = {055006},
          doi = {10.1103/PhysRevD.106.055006},
archivePrefix = {arXiv},
       eprint = {2109.00546},
 primaryClass = {hep-ph},
       adsurl = {https://ui.adsabs.harvard.edu/abs/2022PhRvD.106e5006H}
}

@ARTICLE{Metodiev2017CWoLA,
       author = {{Metodiev}, Eric M. and {Nachman}, Benjamin and {Thaler}, Jesse},
        title = "{Classification without labels: learning from mixed samples in high energy physics}",
      journal = {Journal of High Energy Physics},
         year = 2017,
        month = oct,
       volume = {2017},
       number = {10},
          eid = {174},
        pages = {174},
          doi = {10.1007/JHEP10(2017)174},
archivePrefix = {arXiv},
       eprint = {1708.02949},
 primaryClass = {hep-ph},
       adsurl = {https://ui.adsabs.harvard.edu/abs/2017JHEP...10..174M}
}

@ARTICLE{Nachman2020Anode,
       author = {{Nachman}, Benjamin and {Shih}, David},
        title = "{Anomaly detection with density estimation}",
      journal = {\prd},
         year = 2020,
        month = apr,
       volume = {101},
       number = {7},
          eid = {075042},
        pages = {075042},
          doi = {10.1103/PhysRevD.101.075042},
archivePrefix = {arXiv},
       eprint = {2001.04990},
 primaryClass = {hep-ph},
       adsurl = {https://ui.adsabs.harvard.edu/abs/2020PhRvD.101g5042N}
}

@ARTICLE{Pettee2024StreamCwola,
       author = {{Pettee}, Mariel and {Thanvantri}, Sowmya and {Nachman}, Benjamin and {Shih}, David and {Buckley}, Matthew R. and {Collins}, Jack H.},
        title = "{Weakly supervised anomaly detection in the Milky Way}",
      journal = {\mnras},
         year = 2024,
        month = jan,
       volume = {527},
       number = {3},
        pages = {8459-8474},
          doi = {10.1093/mnras/stad3663},
archivePrefix = {arXiv},
       eprint = {2305.03761},
 primaryClass = {astro-ph.GA},
       adsurl = {https://ui.adsabs.harvard.edu/abs/2024MNRAS.527.8459P}
}

@ARTICLE{Sengupta2025Skycurtains,
       author = {{Sengupta}, Debajyoti and {Mulligan}, Stephen and {Shih}, David and {Raine}, John Andrew and {Golling}, Tobias},
        title = "{SKYCURTAINS: model-agnostic search for stellar streams with Gaia data}",
      journal = {\mnras},
         year = 2025,
        month = jan,
       volume = {536},
       number = {2},
        pages = {1104-1114},
          doi = {10.1093/mnras/stae2570},
archivePrefix = {arXiv},
       eprint = {2405.12131},
 primaryClass = {astro-ph.GA},
       adsurl = {https://ui.adsabs.harvard.edu/abs/2025MNRAS.536.1104S}
}

@ARTICLE{Hallin2025Viamachinae3,
       author = {{Hallin}, Anna and {Shih}, David and {Krause}, Claudius and {Buckley}, Matthew R.},
        title = "{Via Machinae 3.0: A search for stellar streams in Gaia with the CATHODE algorithm}",
      journal = {arXiv e-prints},
         year = 2025,
        month = sep,
          eid = {arXiv:2509.08064},
        pages = {arXiv:2509.08064},
          doi = {10.48550/arXiv.2509.08064},
archivePrefix = {arXiv},
       eprint = {2509.08064},
 primaryClass = {astro-ph.GA},
       adsurl = {https://ui.adsabs.harvard.edu/abs/2025arXiv250908064H}
}

@ARTICLE{Shih2024Viamachinae2,
       author = {{Shih}, David and {Buckley}, Matthew R. and {Necib}, Lina},
        title = "{VIA MACHINAE 2.0: Full-sky, model-agnostic search for stellar streams in Gaia DR2}",
      journal = {\mnras},
         year = 2024,
        month = apr,
       volume = {529},
       number = {4},
        pages = {4745-4767},
          doi = {10.1093/mnras/stae446},
archivePrefix = {arXiv},
       eprint = {2303.01529},
 primaryClass = {astro-ph.GA},
       adsurl = {https://ui.adsabs.harvard.edu/abs/2024MNRAS.529.4745S}
}

@ARTICLE{Shih2022Viamachinae1,
       author = {{Shih}, David and {Buckley}, Matthew R. and {Necib}, Lina and {Tamanas}, John},
        title = "{VIA MACHINAE: Searching for stellar streams using unsupervised machine learning}",
      journal = {\mnras},
         year = 2022,
        month = feb,
       volume = {509},
       number = {4},
        pages = {5992-6007},
          doi = {10.1093/mnras/stab3372},
archivePrefix = {arXiv},
       eprint = {2104.12789},
 primaryClass = {astro-ph.GA},
       adsurl = {https://ui.adsabs.harvard.edu/abs/2022MNRAS.509.5992S}
}

@ARTICLE{Malhan2018Streamfinder,
       author = {{Malhan}, Khyati and {Ibata}, Rodrigo A.},
        title = "{STREAMFINDER - I. A new algorithm for detecting stellar streams}",
      journal = {\mnras},
         year = 2018,
        month = jul,
       volume = {477},
       number = {3},
        pages = {4063-4076},
          doi = {10.1093/mnras/sty912},
archivePrefix = {arXiv},
       eprint = {1804.11338},
 primaryClass = {astro-ph.GA},
       adsurl = {https://ui.adsabs.harvard.edu/abs/2018MNRAS.477.4063M}
}

@ARTICLE{Bonaca2025Streamreview,
       author = {{Bonaca}, Ana and {Price-Whelan}, Adrian M.},
        title = "{Stellar streams in the Gaia era}",
      journal = {\nar},
         year = 2025,
        month = jun,
       volume = {100},
          eid = {101713},
        pages = {101713},
          doi = {10.1016/j.newar.2024.101713},
archivePrefix = {arXiv},
       eprint = {2405.19410},
 primaryClass = {astro-ph.GA},
       adsurl = {https://ui.adsabs.harvard.edu/abs/2025NewAR.10001713B}
}

@ARTICLE{LyndenBell1995Stream,
       author = {{Lynden-Bell}, D. and {Lynden-Bell}, R.~M.},
        title = "{Ghostly streams from the formation of the Galaxy's halo}",
      journal = {\mnras},
         year = 1995,
        month = jul,
       volume = {275},
       number = {2},
        pages = {429-442},
          doi = {10.1093/mnras/275.2.429},
       adsurl = {https://ui.adsabs.harvard.edu/abs/1995MNRAS.275..429L}
}

@ARTICLE{Johnston1998TDE,
       author = {{Johnston}, Kathryn V.},
        title = "{A Prescription for Building the Milky Way's Halo from Disrupted Satellites}",
      journal = {\apj},
         year = 1998,
        month = mar,
       volume = {495},
       number = {1},
        pages = {297-308},
          doi = {10.1086/305273},
archivePrefix = {arXiv},
       eprint = {astro-ph/9710007},
 primaryClass = {astro-ph},
       adsurl = {https://ui.adsabs.harvard.edu/abs/1998ApJ...495..297J}
}

@ARTICLE{Mateu2023Galstreams,
       author = {{Mateu}, Cecilia},
        title = "{galstreams: A library of Milky Way stellar stream footprints and tracks}",
      journal = {\mnras},
         year = 2023,
        month = apr,
       volume = {520},
       number = {4},
        pages = {5225-5258},
          doi = {10.1093/mnras/stad321},
archivePrefix = {arXiv},
       eprint = {2204.10326},
 primaryClass = {astro-ph.GA},
       adsurl = {https://ui.adsabs.harvard.edu/abs/2023MNRAS.520.5225M}
}



\end{document}